\documentstyle[12pt]{article}

\author{Gustav W. Delius
\\ \\
\small{Fakult\"at f\"ur Physik, Universit\"at Bielefeld}\\
\small{Postfach 10 01 31, D-33501 Bielefeld, Germany}\\
\footnotesize{e-mail: delius@physf.uni-bielefeld.de}\\ \\
Yao-Zhong Zhang
\\ \\
\small{Department of Mathematics, University of Queensland}\\
\small{Brisbane Qld 4072, Australia.}\\
\footnotesize{e-mail: yzz@maths.uq.oz.au} }
\title{Finite dimensional representations of\\
quantum affine algebras}

\date{}
\begin{document}

\newcommand{\G}{{\cal G}}
\newcommand{\C}{{\bf C}}

\begin{titlepage}

\maketitle
\begin{abstract}
We give a general construction for finite dimensional representations
of $U_q(\hat{\G})$ where $\hat{\G}$ is a non-twisted affine
Kac-Moody algebra with no derivation and zero central charge.
At $q=1$ this is trivial
because $U(\hat{\G})=U({\G})\otimes \C(x,x^{-1})$ with $\G$
a finite dimensional Lie algebra. But this fact no
longer holds after quantum deformation. In most cases it is necessary to
take the direct sum of several irreducible $U_q({\G})$-modules to form an
irreducible $U_q(\hat{\G})$-module which becomes reducible at $q = 1$.
%This implies that affinizable representations are in general reducible ones.
We illustrate our technique by working out explicit examples for
$\hat{\G}=\hat{C}_2$ and $\hat{\G}=\hat{G}_2$.
These finite dimensional modules determine the multiplet structure
of solitons in affine Toda theory.

\end{abstract}
\vspace{1cm}
hep-th/9403162\\
\end{titlepage}
\newpage

% number equations by section

\newcommand{\sect}[1]{\setcounter{equation}{0}\section{#1}}
\renewcommand{\theequation}{\thesection.\arabic{equation}}

%    set footnote numbering
%
%\renewcommand{\thefootnote}{\fnsymbol{footnote}}
%\setcounter{footnote}{0}

% Put parentheses around equation references

\newcommand{\reff}[1]{eq. (\ref{#1})}

% Shorthands for \begin{equation} and the like

\newcommand{\beq}{\begin{equation}}
\newcommand{\eeq}{\end{equation}}
\newcommand{\bea}{\begin{eqnarray}}
\newcommand{\eea}{\end{eqnarray}}
\newcommand{\bq}{\begin{quote}}
\newcommand{\eq}{\end{quote}}

\newtheorem{Theorem}{Theorem}
\newtheorem{Definition}{Definition}
\newtheorem{Lemma}[Theorem]{Lemma}
\newtheorem{Corollary}[Theorem]{Corollary}
\newcommand{\proof}[1]{{\bf Proof. }
        #1\begin{flushright}$\Box$\end{flushright}}

%This adds space between paragraphs

\parskip 0.3cm

% Some shorthands

\renewcommand{\a}{\alpha}
\renewcommand{\b}{\beta}
\renewcommand{\l}{\lambda}
\newcommand{\s}{\sigma}
\newcommand{\D}{\Delta}
\newcommand{\ug}{$U_q({\G})$ }
\newcommand{\ugd}{$U_q({\G})$}
\newcommand{\ugh}{$U_q(\hat{\G})$ }
\newcommand{\ughd}{$U_q(\hat{\G})$}
\newcommand{\qh}{q^{h_0/2}}
\newcommand{\qmh}{q^{-h_0/2}}
\newcommand{\cq}{$U_q(C_2)$}
\newcommand{\cqh}{$U_q(\hat{C}_2)$}
\newcommand{\gq}{$U_q(G_2)$}
\newcommand{\gqh}{$U_q(\hat{G}_2)$}

\sect{Introduction}

One reason for the importance of quantum algebras \ug in mathematical
physics is their relation to the Yang-Baxter equation: each intertwiner
(R-matrix) for the tensor product of two finite dimensional representations of
a
quantum algebra provides a solution to the quantum Yang-Baxter equation
\cite{Dri85,Jim85,Jim86a,Dri86,Jim86b}.

There are at least two areas where it is important to know solutions
of the spectral parameter {\em dependent} Yang-Baxter equation.
One are integrable
lattice models, where the existence of commuting transfer matrices follows
if the Boltzman weights satisfy the spectral parameter dependent Yang-Baxter
equation. The other are massive integrable quantum field theories where
the spectral parameter dependent Yang-Baxter equation is the consistency
condition of the 2-particle factorization of the scattering matrix. The
spectral parameter in this case is the rapidity of the particles.

The R-matrices $\check{R}_{ab}$ for \ugd, where $\G$ is a finite
dimensional simple Lie algebra, provide solutions to the Yang-Baxter
equation {\em without} a spectral parameter. Here $a,b$ are the labels of
the representation spaces $V_a,\ V_b$, i.e. $\check{R}_{ab}$ is the
intertwiner : $ V_a\otimes V_b\rightarrow V_b\otimes V_a$.
The interesting question is: when can a spectral parameter be introduced
into $\check{R}_{ab}$ so as to obtain a solution of the spectral
parameter {\em dependent} Yang-Baxter equation. The answer is: whenever
$V_a,\ V_b$ carry representations also of the quantum affine algebra
\ughd. Here $\hat{\G}$ is the affine algebra ${\G}\otimes \C(x,x^{-1})$.
The parameter $x$ then consistently provides the spectral
parameter.
\cite{Dri85,Jim85,Jim86a,Dri86,Jim86b}.

This paper is devoted to studying in which cases this affinization
is possible, i.e.
which finite dimensional representations spaces of \ug carry
representations also of \ughd.

For ${\G}=A_n$  all representations are affinizable \cite{Jim86a} (see also
the appendix of \cite{Zha93}).
%This is because in this case the Hecke algebra provides
%a quantum analogue of the symmetric group which can be used to project
%any irrep out of the n-fold tensor product of the fundamental representation.
For other algebras this is not the case. Frenkel \& Reshetikhin \cite{Fre92}
state that ``one generally has to enlarge [an irrep] $V_\lambda$ by
adding certain `smaller' irreducible representations in order to
extend the resulting representation to \ughd. An explicit description of
this extension is an important open problem.'' As far as we know, such a
description is still lacking.

We begin in section \ref{Definition} by defining \ug and \ughd.
Then in section \ref{Example} we give some concrete examples where two irreps
of \ug have to be added together to obtain an irrep of \ughd. The first example
we choose is the 10-dimensional representation of \cq{} which
has to be enlarged by the singlet representation to give an
11-dimensional irrep of \cqh. The second one is the 14-dimensional
representation of \gq{} which again has to be enlarged by the trivial
representation to give a
15-dimensional irrep of \gqh. In section \ref{General} we present our general
procedure for obtaining irreps of \ughd. Our method is based on the
reduction of tensor products of smaller representations. It is
therefore very much in the spirit of the fusion procedure used to construct
rational \cite{Kul81} and trigonometric \cite{Jim86a} R-matrices. The technical
device which we
will use is the tensor product graph \cite{Zha91}. In section \ref{Specific} we
illustrate our general method again in the cases of \cqh{} and \gqh.

Our physical motivation
for this study of finite dimensional representations of quantum affine algebras
comes from the desire to gain a better understanding of the solitons in
affine Toda quantum field theory. These solitons transform in such
representations and we will come back to that point in the discussions
in section \ref{Discussion}.

\sect{Definition of quantum algebras\label{Definition}}

A simple Lie algebra $\G$ is defined through its simple roots
$\a_i,\ i=1\dots r$ by the following relations between its Chevalley
generators $h_i, e_i, f_i,\ i=1\dots r$
\bea\label{classicalrels}
&&[h_i,e_j]=(\a_i,\a_j)e_j,\qquad [h_i,f_j]=-(\a_i,\a_j)f_j,\nonumber\\
&&[e_i,f_j]=\delta_{ij}h_j,\\
%&&\sum_{k=0}^{1-a_{ij}}(-1)^k e_i^{1-a_{ij}-k}e_j e_i^{k}=0
%\qquad(i\neq j)\nonumber\\
%&&\sum_{k=0}^{1-a_{ij}}(-1)^k f_i^{1-a_{ij}-k}f_j f_i^{k}=0
%\qquad(i\neq j)\nonumber
&&(\mbox{ad} e_i)^{1-a_{ij}}e_j=0,\qquad
(\mbox{ad} f_i)^{1-a_{ij}}f_j=0,\qquad (i\neq j)\nonumber
\eea
where $ a_{ij}=2(\a_i,\a_j)/(\a_i,\a_i)$.
The universal enveloping algebra $U({\G})$ is the algebra generated
freely by the Chevalley generators modulo the relations
\reff{classicalrels}.  The quantum algebra \ug is a deformation of
this \cite{Dri85,Jim85} where \reff{classicalrels} is modified to
\bea\label{quantumrels}
&&[h_i,e_j]=(\a_i,\a_j)e_j,\qquad [h_i,f_j]=-(\a_i,\a_j)f_j,\nonumber\\
&&[e_i,f_j]=\delta_{ij}[h_j]_q,\\
%&&\sum_{k=0}^{1-a_{ij}}(-1)^k e_i^{(1-a_{ij}-k)}e_j e_i^{(k)}=0
%\qquad(i\neq j)\nonumber\\
%&&\sum_{k=0}^{1-a_{ij}}(-1)^k f_i^{(1-a_{ij}-k)}f_j f_i^{(k)}=0
%\qquad(i\neq j)\nonumber
&&(\mbox{ad}_q e_i)^{1-a_{ij}}e_j=0,\qquad
(\mbox{ad}_q f_i)^{1-a_{ij}}f_j=0,\qquad (i\neq j)\nonumber
\eea
We have introduced the notation
\beq
[u]_q={q^u-q^{-u}\over q-q^{-1}},~~~
%e_i^{(k)}=\frac{e_i^k}{[k]_{q_i}!},~~~
%f_i^{(k)}=\frac{f_i^k}{[k]_q!},~~~ q_i=q^{(\a_i,\a_i)/2},~~~
%[k]_q!=\prod_{i=1}^k\,[i]_q
\eeq
and $\mbox{ad}_q$ is a q-commutator
\beq
(\mbox{ad}_q e_i)e_j\equiv [e_i,e_j]_q\equiv e_i e_j-q^{(\a_i,\a_j)}e_j e_i.
\eeq
The most important feature of this deformation is that it is still a
Hopf-algebra. The deformed comultiplication is
\bea\label{comultiplication}
\D(h_i)&=&h_i\otimes 1+1\otimes h_i,\nonumber\\
\D(e_i)&=&e_i\otimes q^{h_i/2}+q^{-h_i/2}\otimes e_i,\\
\D(f_i)&=&f_i\otimes q^{h_i/2}+q^{-h_i/2}\otimes f_i.\nonumber
\eea

The finite dimensional representations of \ug have been studied by Rosso
\cite{Ros88} and Lusztig \cite{Lus88}. They found that, for $q$ not a root of
unity, the representation
theory of \ug is exactly analogous to that of $\G$. Each finite-dimensional
irreducible $\G$-module also carries an irrep of \ug and the irreps of
\ug are simply deformations of those of $\G$.

Non-twisted affine Lie algebras $\tilde{\G}$, as defined by Kac
\cite{Kac90}, can be realized as
$\tilde{\G}={\G}\otimes\C(x,x^{-1})\oplus \C c\oplus \C d$,
where $\C(x,x^{-1})$ is the algebra of Laurent polynomials in $x$, $c$
is a central charge and $d$ is a derivation. In this paper we are only
interested in the algebra $\hat{\G}={\G}\otimes\C(x,x^{-1})\oplus \C c$
obtained from $\tilde{\G}$ by dropping the derivation.  The algebra with
derivation does not have finite dimensional representations. Following
a widespread custom in the literature we will call also the algebra
$\hat{\G}$ an affine algebra.
{}From a finite dimensional representation $\pi$ of $\hat{\G}$ one can easily
obtain a loop representation of the algebra with derivation $\tilde{\G}$.

To generate the affine algebra $\hat{\G}$
it is sufficient to add one more pair of raising and lowering operators
and one more Cartan subalgebra generator to the Chevalley basis, namely
\beq\label{classicale0}
e_0=f_\psi\otimes x,\qquad f_0=e_\psi\otimes x^{-1},
\qquad h_0=(c-h_\psi)\otimes 1,
\eeq
where $\psi$ is the highest root of $\G$ and $e_\psi, f_\psi$ are the
corresponding raising and lowering operators \cite{Kac90}.
The new Chevalley generators again satisfy the relations
\reff{classicalrels}, this time with $i,j=0\dots r$ and $\a_0=-\psi$.
The central charge $c$ will play no
role in this paper because it is represented as zero on all finite
dimensional modules.

The {\it quantum} affine algebra \ugh is defined analogously by the
relations \reff{quantumrels}.
There is one important difference between $U(\hat{\G})$ and \ugh, i.e.
between the classical and the quantum case. Classically
$e_0$ and $f_0$ are elements of $U({\G})\otimes \C(x,x^{-1})$,
see \reff{classicale0}, and thus
\beq\label{loop}
U(\hat{\G})
=U({\G})\otimes \C(x,x^{-1})\oplus \C c.
\eeq
In the quantum case however,
generically $e_0$ and $f_0$ are not elements of $U_q({\G})\otimes
\C(x,x^{-1})$,
as will be seen in the next section. Thus
\beq
U_q(\hat{\G})
\neq U_q({\G})\otimes \C(x,x^{-1})\oplus \C c.
\eeq
The only known exceptions to this are ${\G}=A_n$ \cite{Jim86a} (see also
the appendix of \cite{Zha93} for details).

Because of \reff{loop}, any $U({\G})$-module is also a
$U(\hat{\G})$-module on which $x$ and $c$ are represented trivially.
This is no longer true in the quantum case.
Some representations spaces of \ug may not carry a representation of $e_0$
and $f_0$. Obviously those and only those representations spaces which carry
a representation of $e_0$ and $f_0$ carry a representation of \ughd.
It is the aim of this paper to construct such representations.

\sect{Examples of representations\label{Example}}

The easiest way to prove that $e_0$ is not an element of $U_q({\G})
\otimes \C(x,x^{-1})$ in general is to give some simple examples of
representation spaces of
\ug which do not carry a representation of $e_0$ and $f_0$. As we will see, one
usually has to take a direct sum of two (or more) irreps of \ug to form
an irrep of \ughd.

The first simple example is the 10-dimensional irrep of \cq.
$C_2$ has two simple roots $\a_1$ and $\a_2$ which satisfy
$2(\a_1,\a_1)$ = $(\a_2,\a_2)$ =
$-2(\a_1,\a_2)$ = $-2(\a_2,\a_1)=2$. The 10-dimensional representation is the
adjoint representation and its weights are the roots
$\{2\a_1+\a_2\,, \,\a_1+\a_2\,, \,\a_1\,, \,\a_2\,, \,0\,, \,
0\, ,\,-\a_2\,, \,-\a_1\,,
\newline
-\a_1-\a_2\,, \,-2\a_1-\a_2\}$.
The matrix forms
of $h_1,~h_2,~e_1$ and $e_2$ in this representation are
\bea\label{rep10}
&&h_1=e_{11}-e_{33}+e_{44}-e_{77}+e_{88}-e_{10,10}\nonumber\\
&&h_2=e_{22}+2e_{33}-e_{44}+e_{77}-2e_{88}-e_{99}\nonumber\\
&&e_1=f_1^{\rm t}=e_{12}+e_{23}+e_{45}+e_{57}+e_{89}+e_{9,10}\\
&&e_2=f_2^{\rm t}=e_{24}+e_{35}+e_{58}+e_{79}+([2]_q-1)^{1/2}
  (e_{36}+e_{68})\nonumber
\eea
where ${\rm t}$ stands for transpose and $e_{ij}$ is the matrix with
$1$ in entry $i,j$ and $0$ elsewhere.

One would now like to find two other matrices $e_0$ and $f_0$ which
satisfy the defining relations \reff{quantumrels}. One can make a general
Ansatz
and then at first impose all relations except $[e_0,f_0]=[h_0]_q$ and the
q-Serre relation involving $e_0$ and $f_0$. At
this point one finds that $e_0$ and $f_0$ are already completely
determined up to an overall constant. Unfortunately they do {\em not}
satisfy $[e_0,f_0]=[h_0]_q$ and the q-Serre relations, and this shows that
this irrep of \cq{} can not be extended to a representation of \cqh.

Next we consider a direct sum of the 10-dimensional irrep with the trivial
one-dimensional representation. For this 11-dimensional reducible
representation of \cq, the matrix
form for $h_1,~h_2,~e_1$ and $e_2$ looks the same as above. Now it is
possible to find  matrices $e_0$ and $f_0$ satisfying all of the relations
\reff{quantumrels}:
\bea\label{e011}
e_0=f_0^{\rm t}&=&e_{51}+e_{72}+e_{94}+e_{10,5}-([2]_q-1)^{-1/2}
   (e_{61}+e_{10,6})+\nonumber\\
& &+[2]_q^{1/2}\left (\frac{[2]_q-2}{[2]_q-1}\right
)^{1/2}(e_{10,11}+e_{11,1}).
\eea
This representation of \cqh{} is seen to be irreducible. It becomes
reducible at $q=1$, as can be seen from the coefficient of the last term.

The second example we want to give is the 14-dimensional irrep of \gq.
The simple roots of $G_2$ are $\a_1$ and $\a_2$ which satisfy
$(\a_1,\a_1)=3(\a_2,\a_2)=-2(\a_1,\a_2)=-2(\a_2,\a_1)=6$. The 14-dimensional
representation is the
adjoint representation with weights equal to the roots
$\{2\a_1+3\a_2\,, \,\a_1+3\a_2\,, \,\a_1+2\a_2\,, \,\a_1+\a_2\,, \,\a_1\,,
\a_2\,,\, 0\,,
0\,, \,-\a_2\,, \,-\a_1\,, \,-\a_1-\a_2\,, \,-\a_1-2\a_2\,, \,
-\a_1-3\a_2\,, \,-2\a_1-3\a_2\}$.
The matrix forms of $h_1,~h_2,~e_1$ and $e_2$ for the 14-dimensional
irrep of \gq{} are
\bea\label{rep14}
h_1&=&3e_{11}-3e_{22}+3e_{44}+6e_{55}-3e_{66}+3e_{99}\nonumber\\
   & &-6e_{10,10}-3e_{11,11}+3e_{13,13}-3e_{14,14}\nonumber\\
h_2&=&3e_{22}+e_{33}-e_{44}-3e_{55}+2e_{66}-2e_{99}\nonumber\\
   & &+3e_{10,10}+e_{11,11}-e_{12,12}-3e_{13,13}\nonumber\\
e_1&=&f_1^{\rm t}\ =[3]_q^{1/2}(e_{12}+e_{46}+e_{9,11}+e_{13,14})
   +\frac{[3]_q}{[2]_q^{1/2}}
     (e_{58}+e_{8,10})\nonumber\\
   & &+[3]_q^{1/2}\left (\frac{[6]_q}{[3]_q}-\frac{[3]_q}{[2]_q}\right
      )^{1/2}(e_{57}+e_{7,10})\nonumber\\
e_2&=&f_2^{\rm t}\ =[3]_q^{1/2}(e_{23}+e_{45}+e_{10,11}+e_{12,13})+\nonumber\\
   & &+([3]_q+1)^{1/2}(e_{34}+e_{11,12})+[2]_q^{1/2}(e_{68}+e_{89})\label{g2}
\eea
Again it can be shown that this irrep of \gq{} can not be extended to a
representation of \gqh. Next we consider the direct sum of this
irrep with the trivial
representation of \gq. Obviously the matrix form of $h_1,~h_2,e_1$ and $e_2$
for this 15-dimensional reducible representation  are the
same as (\ref{g2}). It turns out that this reducible representation can
be extended to an irrep of \gqh. The explicit expressions
for $e_0$ and $f_0$ are:
\bea\label{e015}
e_0&=&f_0^{\rm t}\ =[3]_q^{1/2}\left (\frac{[6]_q}{[3]_q}-\frac{[3]_q}{[2]_q}
      \right )^{-1/2}(e_{71}+e_{14,7})+[3]_q^{1/2}(e_{10,2}
      +e_{11,3}+e_{12,4}+\nonumber\\
   & &+e_{13,5})+\left (\frac{[6]_q^2[2]_q-[3]_q^2[6]_q-[3]_q^2[2]_q}
     {[6]_q[2]_q-[3]_q^2}\right )^{1/2}(e_{14,15}+e_{15,1})
\eea
which defines a 15-dimensional irrep of \gqh. This irrep becomes reducible
only when $q=1$.

\sect{General construction\label{General}}

Because $e_0$ does not exist as an element in $U_q({\G})\otimes
\C(x,x^{-1})$, we will have to construct
$\pi(e_0)$ for each representation $\pi$ separately.
Clearly we can not proceed as
in the previous section but need a general construction.

Let $V_\l$ be an irreducible finite dimensional
$\G$-module and $\pi_\l :U_q({\G})\rightarrow
\mbox{End}(V_\l)$ the representation of \ug which it carries. Assume that on
this module it is possible to define $\pi_\l(e_0)$ and $\pi_\l(f_0)$ and thus
make it into an irreducible representation of \ughd.{}~
\footnote{Examples are the ``undeformed'' representations for
$U_q({\G})$ with ${\G}=A_n,~ B_n,~ C_n,~ D_n,~ E_6$ and $E_7$, which
are affinizable, that is $e_0$ and $f_0$ exist for those representations,
and the minimal representations for $U_q(G_2)$ and $U_q(F_4)$, which are
deformed representations but nevertheless are affinizable.}
We start with this irrep and want to construct, using it, further irreps
of \ughd. To this end we
look at the tensor product $V_\l\otimes V_\l$ which carries the
\ugd-representation
\beq
\Pi(g)=(\pi_\l\otimes\pi_\l)\D(g),\qquad\qquad g\in U_q({\G}).
\eeq
It is a reducible representation of \ug and it is known that the
decomposition into irreps is the same as in the classical case
\cite{Ros88,Lus88}
\beq\label{decomposition}
V_\l\otimes V_\l=\bigoplus_\mu V_\mu.
\eeq
We want to see on which of these irreducible modules $V_\mu$ or on which
direct sums of them we can define  irreps of \ughd. It can be checked that
the following  defines a representation of \ugh on $V_\l\otimes V_\l$
\bea
\Pi^{a}(g)&=&\Pi(g),\qquad\qquad g\in U_q({\G}),\\
\label{e0rep}
\Pi^{a}(e_0)&=&(\pi_\l\otimes\pi_\l)(e_0\otimes q^{h_0/2}
        +a q^{-h_0/2}\otimes e_0),\\
\label{f0rep}
\Pi^{a}(f_0)&=&(\pi_\l\otimes\pi_\l)(f_0\otimes q^{h_0/2}
        +a^{-1} q^{-h_0/2}\otimes f_0),\\
\Pi^{a}(h_0)&=&(\pi_\l\otimes\pi_\l)(h_0\otimes 1+1\otimes h_0),
\eea
for any choice of the parameter $a\in \C$.
We will see that for generic value of $a$ the representation $\Pi^{a}$ is
irreducible but that it becomes reducible for special values and at these
values we can define irreducible representations on submodules of
$V_\l\otimes V_\l$.

To visualize the reducibility of the representation $\Pi^{a}$
we describe it by a directed graph.
A similar graph, called the tensor product graph, was first introduced
in \cite{Zha91} and we will rely heavily on ideas from that paper.

\begin{Definition}\label{reducibilitygraph}
The {\bf reducibility graph} $G^a$ associated to the representation $\Pi^{a}$
of \ugh is a directed graph whose vertices are the irreducible $\G$-modules
$V_\mu$ appearing in the decomposition \reff{decomposition} of
$V_\l\otimes V_\l$. There is an edge directed from a vertex $V_\nu$
to a vertex $V_\mu$ iff
\beq\label{link}
P_\mu\Pi^{a}(e_0) P_\nu\neq 0\qquad\mbox{or}\qquad
P_\mu\Pi^{a}(f_0) P_\nu\neq 0,
\eeq
where $P_\mu$ is the projector from $V_\l\otimes V_\l$ onto $V_\mu$.
\end{Definition}

\begin{figure}
\unitlength=1mm
\linethickness{0.4pt}
\begin{picture}(122.60,25.00)
\put(30.00,20.00){\circle*{5.20}}
\put(60.00,20.00){\circle*{5.20}}
\put(90.00,20.00){\circle*{5.20}}
\put(120.00,20.00){\circle*{5.20}}
\put(45.00,22.50){\oval(30.00,5.00)[t]}
\put(45.00,17.50){\oval(30.00,5.00)[b]}
\put(105.00,22.50){\oval(30.00,5.00)[t]}
\put(105.00,17.50){\oval(30.00,5.00)[b]}
\put(75.00,20.00){\oval(30.00,0.00)[t]}
\put(30.00,7.00){\makebox(0,0)[cc]{1}}
\put(60.00,7.00){\makebox(0,0)[cc]{14}}
\put(90.00,7.00){\makebox(0,0)[cc]{27}}
\put(120.00,7.00){\makebox(0,0)[cc]{7}}
\put(40.00,25.00){\vector(1,0){5.00}}
\put(50.00,15.00){\vector(-1,0){5.00}}
\put(80.00,20.00){\vector(-1,0){5.00}}
\put(100.00,25.00){\vector(1,0){5.00}}
\put(110.00,15.00){\vector(-1,0){5.00}}
\end{picture}
\caption{The reducibility graph $G^a$ for the $7\otimes 7$ of \gq{}
at $a=q^{2}$. The vertices are labeled by the dimension of the corresponding
irrep of \gq.\label{redgraph}}
\end{figure}
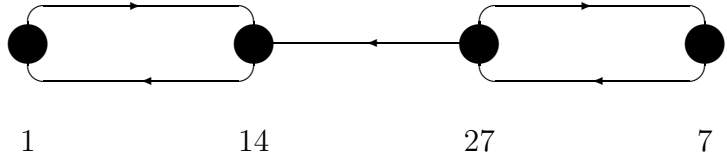

For an example of a reducibility graph see figure \ref{redgraph}.
According to the definition, an arrow from $V_\nu$ to $V_\mu$ indicates
that $\Pi^{a}(e_0)$ or $\Pi^{a}(f_0)$
can bring us from the module $V_\nu$ to the module $V_\mu$.
It implies that a \ughd -submodule of $V_\l\otimes V_\l$ which contains
$V_\nu$ also has to contain $V_\mu$. In other words, the
\ughd -submodules are described by those subgraphs from which
no arrows point outside that subgraph. We formulate this in the next
definition and theorem.

\begin{Definition}\label{connected}\label{closed}
A subgraph $G^\prime$ of a graph $G$ is called
\begin{itemize}
\item{{\bf two-way connected} if for any pair $V_\mu, V_\nu$ of vertices
in $G^\prime$ there exist \underline{directed} paths from $V_\mu$ to $V_\nu$
\underline{and} from $V_\nu$ to $V_\mu$. }
\item{{\bf simply two-way connected} if it is two-way connected and
becomes non-two-way connected if any edge is removed,}
\item{{\bf closed} if there is no edge pointing from any vertex in
$G^\prime$ to a vertex outside $G^\prime$.}
\end{itemize}
\end{Definition}

\begin{Theorem}\label{representation}
Every closed subgraph $G^\prime$ of a reducibility graph $G^a$
defines a representation $(V^\prime,\pi^\prime)$
of \ughd. The representation space $V^\prime$ is the direct sum of the
irreducible $\G$-modules corresponding to the vertices in $G^\prime$
\beq
V^\prime=\bigoplus_{V_\mu\in G^\prime}V_\mu
\eeq
Let $P^\prime$ be the projector from $V_\l\otimes V_\l$ onto $V^\prime$:
$P^\prime=\sum_{V_\mu\in G^\prime}P_{\mu}$.
The representation map $\pi^\prime: U_q(\hat{\G})\rightarrow
\mbox{End}(V^\prime)$ is given by
\beq\label{rep}
\pi^\prime(g)=P^\prime\Pi^a(g)P^\prime\qquad g\in U_q(\hat{\G}).
\eeq
If the subgraph $G^\prime$ is two-way connected then the representation
$\pi^\prime$
is irreducible.
\end{Theorem}
\proof{We only need to show that \reff{rep} defines a representation
using the fact that $\Pi^a$ does. This becomes trivial with the following
observation: Because $G^\prime$ is closed, we know that $\Pi^a(e_0)v^\prime
\in V^\prime$ for all $v^\prime \in V^\prime$ and similarly for $f_0$.
Also $\Pi^a(g)v^\prime\in V^\prime$ for all $g\in U_q(\G)$ because $V^\prime$
is a sum of \ugd-modules. Together this gives that
$\Pi^a(g)v^\prime\in V^\prime$ for all $g\in $\ughd. Combining this
with $P^\prime v^\prime =v^\prime$ we have that $P^\prime\Pi^a(g)
P^\prime=\Pi^a(g) P^\prime$ and therefore the projectors $P^\prime$
can be pulled outside in checking that the relations \reff{quantumrels}
are satisfied.
The irreducibility of $\pi^\prime$ follows immediately from the
two-way connectedness of $G^\prime$.}

Applying this theorem to the example in figure \ref{redgraph} we see that
there is a 15-dimensional irreducible representation of \gqh. This
reproduces the representation which we found in section \ref{Example}.

The usefulness of Theorem \ref{representation} lies in the fact that the
reducibility graph encodes the reducibility of the tensor product and
can in most cases be easily constructed using only elementary Lie-algebra
representation theory. The facts needed for this purpose are contained in the
following lemmas.

\begin{Lemma}\label{connectedness}
The reducibility graph $G^a$ is two-way connected for generic values of $a$. It
can be
non-two-way connected only if
\beq\label{a}
a=q^{C(\mu)-C(\nu)\over 2}
\eeq
where $C(\l)=(\l,\l+2\rho)$ is the value of the quadratic Casimir on $V_\l$.
\end{Lemma}

\proof{Here we can follow \cite{Zha91}, who defined a similar graph.
For clarity we first consider the classical case $q=1$.
To make the notation simpler we will from now on
drop the $\pi_\l$ and write simply $e_i$ instead of $\pi_\l(e_i)$ etc.
$V_\l$ was irreducible by definition. This means that by repeatedly acting
with the tensor operators $T=\{g\otimes 1|g\in{\G}\}$ and $\bar{T}
=\{1\otimes g|g\in{\G}\}$ we can obtain any vector in $V_\l\otimes V_\l$
from any other. In particular these tensor operators connect together
all irreducible $\G$-modules $V_\mu$ contained in $V_\l\otimes V_\l$.
Now $e_0\otimes 1$ and $1\otimes e_0$ are just the lowest
components of these tensor operators (because at $q=1\ e_0=f_\psi$)
and therefore also connect together all modules $V_\mu$. Furthermore
$e_0\otimes 1$ by itself or a linear combination of $e_0\otimes 1$
with $1\otimes e_0$ will suffice because of the proportionality
\beq\label{prop}
P_\mu(e_0\otimes 1)P_\nu=-P_\mu(1\otimes e_0)P_\nu\qquad\qquad
(\mbox{at }q=1),
\eeq
which follows from the fact that $P_\mu$ commutes with
$e_0\otimes 1+1\otimes e_0$. This shows for $q=1$ that
$\Pi^{a}(e_0)=e_0\otimes 1+a\, 1\otimes e_0$ connects all irreps in
$V_\l\otimes V_\l$ unless $a=1$. At $a=1$ $P_\mu\Pi^{a}(e_0) P_\nu$
is always zero according to \reff{prop}. Exactly the same can be
said about $f_0$.
Thus at $q=1$ $G^a$ is two-way connected except at $a=1$ where
it is completely disconnected. This complete disconnectedness at $a=1$
implies according to Lemma \ref{representation} that every irreducible
$\G$-module appearing in the tensor product carries a representation
of $U(\hat{\G})$, which we observed already in section 2.

The fact that the reducibility graph is two-way connected for generic values of
$a$
in the classical case $q=1$ implies that this is also true in the quantum
case $q\neq 1$. This is so because an edge which is present at $q=1$
can not be absent for $q\neq 1$. Otherwise $P_\mu\Pi^{a}(e_0)P_\nu$ would not
have a smooth limit as $q\rightarrow 1$.
This proves the first statement of the theorem.

To determine the non-generic values of $a$ at which the reducibility
graph may be non-two-way connected we use the quantum analogue of \reff{prop}:
\bea\label{qprop}
q^{-C(\mu)/2}\epsilon_\mu P_\mu(e_0\otimes\qh)P_\nu&=&
q^{-C(\nu)/2}\epsilon_\nu P_\mu(\qmh\otimes e_0)P_\nu,\nonumber\\
q^{-C(\mu)/2}\epsilon_\mu P_\mu(\qmh\otimes f_0)P_\nu&=&
q^{-C(\nu)/2}\epsilon_\nu P_\mu(f_0\otimes\qh)P_\nu,
\eea
where $\epsilon_\mu$ is the parity of the representation $V_\mu$ in
$V_\l\otimes V_\l$.
To derive \reff{qprop} consider the $\check{R}$-matrix on $V_\l\otimes V_\l$.
According to Jimbo it is determined by the equations \cite{Jim85}
\bea\label{jimbo}
&&[\check{R}(x),\Pi(\D(a))]=0\qquad\forall a\in U_q({\G}),\\
&&\check{R}(x)\left(x e_0\otimes\qh+\qmh\otimes e_0\right)=
\left(e_0\otimes\qh+\qmh\otimes x e_0\right)\check{R}(x),\nonumber\\
&&\check{R}(x)\left(x^{-1} f_0\otimes\qh+\qmh\otimes f_0\right)=
\left(f_0\otimes\qh+\qmh\otimes x^{-1} f_0\right)\check{R}(x),\nonumber
\eea
In the limit $x\rightarrow\infty$ one obtains the spectral parameter
independent $\check{R}$-matrix and \reff{jimbo} reduces to
\bea
[\check{R},\Pi(\D(a))]=0&&\forall a\in U_q({\G}),\\
\label{braide0}
\check{R}\left(e_0\otimes\qh\right)&=&
\left(\qmh\otimes e_0\right)\check{R},\\
\label{braidf0}
\check{R}\left(\qmh\otimes f_0\right)&=&
\left(f_0\otimes \qh\right)\check{R},
\eea
$\check{R}$ is given by the formula
\beq\label{braidmatrix}
\check{R}=q^{C(\l)}\sum_\mu q^{-C(\mu)/2} \epsilon_\mu P_\mu.
\eeq
This was proved in the case where $V_\l\otimes V_\l$ is multiplicity free
by Reshetikhin \cite{Res87}~\footnote{Our $\check{R}$ is the inverse of
the R-matrix in this reference.}
and in the general case by Gould \cite{Gou92}.
By inserting \reff{braidmatrix} into \reff{braide0} and \reff{braidf0}
and multiplying by $P_\mu$
from the left and by $P_\nu$ from the right we obtain equations
\reff{qprop}.

Comparison of \reff{qprop} with \reff{e0rep} and \reff{f0rep}
immediately provides the second statement of the theorem, provided
$\epsilon_\mu\epsilon_\nu=-1$. To see this later fact we observe
that the permutation matrix $\sigma$ (defined by $\sigma(v\otimes
v^\prime)=(v^\prime\otimes v)$) satisfies $P_\mu\sigma=\sigma P_\mu=
\epsilon_\mu P_\mu$ and thus
\beq
P_\mu(e_0\otimes 1)P_\nu=P_\mu\sigma(1\otimes e_0)\sigma P_\nu=
\epsilon_\mu\epsilon_\nu P_\mu(1\otimes e_0)P_\nu.
\eeq
Comparison of this with \reff{prop} gives $\epsilon_\mu\epsilon_\nu=-1$.
}

\begin{Lemma}\label{c}
\bea
P_\mu\Pi^a(e_0)P_\nu\neq 0\qquad &\Rightarrow &\qquad
V_\mu\subset V_{\mbox{adj}}\otimes V_\nu,\label{ce0}\\
P_\mu\Pi^a(f_0)P_\nu\neq 0\qquad &\Rightarrow &\qquad
V_\mu\subset V_{\mbox{adj}}\otimes V_\nu.\label{cf0}
\eea
\end{Lemma}
\proof{The proof uses the concept of tensor operators. These are well
explained in appendix B of \cite{Zha91}. There it is also shown that
$X=\qmh e_0\otimes 1$ is the lowest component of an adjoint tensor
operator\footnote{In comparing with reference \cite{Zha91} one should
replace $q$ by $q^{-1}$ because \cite{Zha91} uses the opposite coproduct.}.
This implies that the vector $X v_\nu$ for $v_\nu\in V_\nu$
must lie in a representation $V_\mu$ which is contained in
$V_{\mbox{adj}}\otimes V_\nu$. The same therefore is true for
$(e_0\otimes\qh)v_\nu=\Delta{(\qh)}X v_\nu$. Similarly also
$\bar{X}=1\otimes\qh e_0$ is the lowest component of an adjoint tensor
operator and thus also $(\qmh\otimes e_0)v_\nu$ must lie in a
representation which is contained in $V_{\mbox{adj}}\otimes V_\nu$.
This is therefore also true for $\Pi^a(e_0)v_\nu$ and \reff{ce0}
follows.
Using similarly that $\qh f_0\otimes q^{h_0}$ and $q^{-h_0}
\otimes \qh f_0$ are the highest components of adjoint tensor
operators one shows \reff{cf0}.}
This lemma prompts us to define another
directed graph associated to the tensor product $V_\l\otimes V_\l$.
\begin{Definition}\label{tpg}
The {\bf tensor product graph} $\Gamma$ associated to a tensor
product $V_\l\otimes V_\l$ is a directed graph whose vertices are the
irreducible $\G$-modules appearing in the decomposition
\reff{decomposition} of $V_\l\otimes V_\l$.
There is an edge directed from vertex $V_\nu$
to a vertex $V_\mu$ iff
\beq
V_\mu\subset V_{\mbox{adj}}\otimes V_\nu.
\eeq
\end{Definition}
Combining Lemma \ref{c} and Lemma \ref{connectedness} we arrive at
\begin{Lemma}\label{t}
If the tensor product graph $\Gamma$ is simply two-way connected then it is
equal to
the reducibility graph $G^a$ for generic value of $a$.
\end{Lemma}
\proof{According to Lemma \ref{c} the tensor product graph contains
every edge that is contained in the reducibility graph. If the tensor product
graph is simply two-way connected than the reducibility graph has to contain
all its
edges, otherwise it could no longer be two-way connected and would violate
Lemma
\ref{connectedness}.}

Lemma \ref{t} is very useful in constructing reducibility graphs because it
is easy to construct tensor product graphs. Many worked out examples of
undirected tensor product graphs can be found in \cite{Zha91}.
To obtain the
directed tensor product graph as defined in Definition \ref{tpg} from
the undirected graphs in \cite{Zha91} one has to replace every undirected
edge by two directed edges in opposite directions. All known examples
are simply two-way connected and so the Lemma \ref{t} applies.
We do not yet know exactly how general this simply connectedness of
tensor product graphs is. But even when the tensor product graph is
multiply connected we still have the following theorem:

\begin{Theorem}\label{theorem2}
Let $V_\l$ be an irreducible $\G$-module which carries a representation
$\pi_\l$ of \ughd. Let $\Gamma$ be the tensor product graph associated
with $V_\l\otimes V_\l$.
Let $G^\prime$ be any simply two-way connected subgraph of $\Gamma$
which can be made closed by deleting just one directed edge from $\Gamma$.
Let $V_\nu$ be the origin and $V_\mu$ be the destination of this edge.
Let $a=q^{\left(C(\mu)-C(\nu)\right)/2}$.

Then $G^\prime$ is a closed two-way connected subgraph of the reducibility
graph
$G^a$ and carries an irreducible representation of \ugh as in Theorem
\ref{representation}.
\end{Theorem}
\proof{The proof that $G^\prime$ is a subgraph of the generic reducibility
graph is similar to the proof of Lemma \ref{t}. The reason why it is a
subgraph of $G^a$ for the particular $a$ is that according to Lemma
\ref{connectedness} at this $a$ the reducibility graph looses the edge
directed from $V_\nu$ to $V_\mu$. $G^\prime$ defines an irreducible
representation because Theorem \ref{representation} applies.}

Theorem \ref{theorem2} is very easy to apply in practice and we will
demonstrate its use in the next section.

\begin{figure}
\unitlength=1.00mm
\linethickness{0.4pt}
\begin{picture}(105.60,25.00)
\put(43.00,15.00){\circle*{5.20}}
\put(73.00,15.00){\circle*{5.20}}
\put(103.00,15.00){\circle*{5.20}}
\put(58.00,17.50){\oval(30.00,5.00)[t]}
\put(58.00,12.50){\oval(30.00,5.00)[b]}
\put(88.00,17.50){\oval(30.00,5.00)[t]}
\put(88.00,12.50){\oval(30.00,5.00)[b]}
\put(43.00,2.00){\makebox(0,0)[cc]{5}}
\put(73.00,2.00){\makebox(0,0)[cc]{10}}
\put(103.00,2.00){\makebox(0,0)[cc]{1}}
\put(53.00,20.00){\vector(1,0){5.00}}
\put(63.00,10.00){\vector(-1,0){5.00}}
\put(83.00,20.00){\vector(1,0){5.00}}
\put(93.00,10.00){\vector(-1,0){5.00}}
\put(58.00,25.00){\makebox(0,0)[cc]{$q^{1}$}}
\put(58.00,4.00){\makebox(0,0)[cc]{$q^{-1}$}}
\put(88.00,4.00){\makebox(0,0)[cc]{$q^{3}$}}
\put(88.00,25.00){\makebox(0,0)[cc]{$q^{-3}$}}
\end{picture}
\caption{The tensor product graph for the $4\otimes 4$ of \cq.
\label{c2four}}
\end{figure}
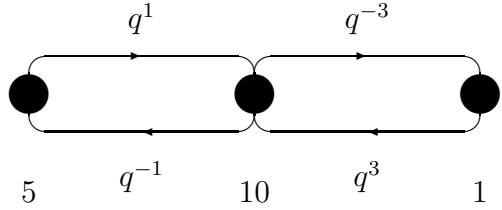

\sect{Specialization to \cqh{} and \gqh\label{Specific}}
(i) \cq:\\
The fundamental 4-dimensional irrep of \cq{} is undeformed and can be
extended to an irrep of \cqh. We will use Theorem \ref{theorem2} to
construct further irreps of \cqh{} from the tensor product\footnote{
We denote the \cq-irreps by their dimension.}
\beq
4\otimes 4=10\bigoplus 5\bigoplus 1
\eeq
The associated tensor product graph is shown in figure \ref{c2four}.
Because it is simply two-way connected it gives also the generic reducibility
graph. The numbers associated to the edges in figure \ref{c2four}
are the values of $a$ from Theorem \ref{theorem2}, i.e., the values
at which the edge disappears from the reducibility graph. They are
determined, using \reff{a}, from $C(1)=0, C(5)=4, C(10)=6$.

We read off from the graph that $\pi^a$ defines a 5-dimensional irrep
of \cqh{} at $a=q$, a 1-dimensional irrep at $a=q^{3}$, a
$(10+5)=15$-dimensional irrep at $a=q^{-3}$ and a $10+1=11$-dimensional
irrep at $a=q^{-1}$, besides of course the $10+5+1=16$-dimensional irrep
at generic $a$. We note in particular that, because the 10-dimensional
irrep of \cq{} appears in the middle of the tensor product graph, there
is no possibility of having the 10-dimensional irrep in a closed
component by itself and thus no irrep of \cqh{} can be defined on it
by itself. The 10 has to be enlarged by adding either the 1, the 5,
or both, before it carries a representation of \cqh. This reproduces
our observation from section \ref{Example}.

We can also derive the representation matrices \reff{rep10}, \reff{e011}
from the general expression \reff{rep}. For this we only need to determine
the q-Clebsch-Gordan coefficients of \cq{} for the decomposition of
the $4\otimes 4$. These can easily be calculated by elementary methods.
We did this using Mathematica.

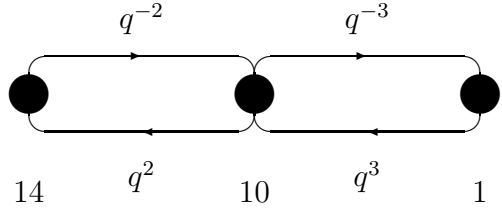
\begin{figure}
\unitlength=1.00mm
\linethickness{0.4pt}
\begin{picture}(105.60,25.00)
\put(43.00,15.00){\circle*{5.20}}
\put(73.00,15.00){\circle*{5.20}}
\put(103.00,15.00){\circle*{5.20}}
\put(58.00,17.50){\oval(30.00,5.00)[t]}
\put(58.00,12.50){\oval(30.00,5.00)[b]}
\put(88.00,17.50){\oval(30.00,5.00)[t]}
\put(88.00,12.50){\oval(30.00,5.00)[b]}
\put(43.00,2.00){\makebox(0,0)[cc]{14}}
\put(73.00,2.00){\makebox(0,0)[cc]{10}}
\put(103.00,2.00){\makebox(0,0)[cc]{1}}
\put(53.00,20.00){\vector(1,0){5.00}}
\put(63.00,10.00){\vector(-1,0){5.00}}
\put(83.00,20.00){\vector(1,0){5.00}}
\put(93.00,10.00){\vector(-1,0){5.00}}
\put(58.00,25.00){\makebox(0,0)[cc]{$q^{-2}$}}
\put(58.00,4.00){\makebox(0,0)[cc]{$q^{2}$}}
\put(88.00,4.00){\makebox(0,0)[cc]{$q^{3}$}}
\put(88.00,25.00){\makebox(0,0)[cc]{$q^{-3}$}}
\end{picture}
\caption{The tensor product graph for the $5\otimes 5$ of \cq.
\label{c2five}}
\end{figure}

Contrary to the 10-dimensional irrep of \cq, the 5-dimensional irrep
can carry an irrep of \cqh{} by itself. We can repeat the above
analysis for the tensor product
\beq
5\otimes 5=14\bigoplus 10\bigoplus 1.
\eeq
The associated tensor product graph is shown in figure \ref{c2five}.
The truncation values shown above the edges are determined from the
Casimir values given earlier and $C(14)=10$.

We see from the graph that the 14-dimensional irrep of \cq{} can carry
an irrep of \cqh, but that again the 10-dimensional irrep of \cq{}
needs to be extended, either by the 1 or by the 14, again reaffirming
our observation from section \ref{Example}. We may continue the above procedure
using the 14-dimensional irrep and get higher irreps of \cqh.

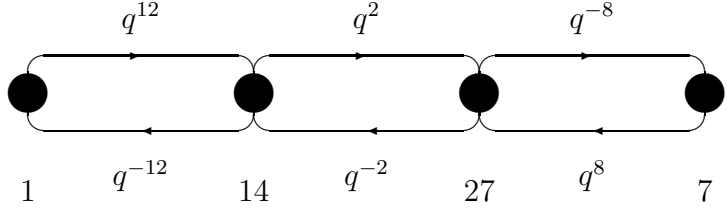
\begin{figure}
\unitlength=1.00mm
\linethickness{0.4pt}
\begin{picture}(122.60,25.00)
\put(30.00,15.00){\circle*{5.20}}
\put(60.00,15.00){\circle*{5.20}}
\put(90.00,15.00){\circle*{5.20}}
\put(120.00,15.00){\circle*{5.20}}
\put(45.00,17.50){\oval(30.00,5.00)[t]}
\put(45.00,12.50){\oval(30.00,5.00)[b]}
\put(75.00,17.50){\oval(30.00,5.00)[t]}
\put(75.00,12.50){\oval(30.00,5.00)[b]}
\put(105.00,17.50){\oval(30.00,5.00)[t]}
\put(105.00,12.50){\oval(30.00,5.00)[b]}
\put(30.00,2.00){\makebox(0,0)[cc]{1}}
\put(60.00,2.00){\makebox(0,0)[cc]{14}}
\put(90.00,2.00){\makebox(0,0)[cc]{27}}
\put(120.00,2.00){\makebox(0,0)[cc]{7}}
\put(40.00,20.00){\vector(1,0){5.00}}
\put(50.00,10.00){\vector(-1,0){5.00}}
\put(70.00,20.00){\vector(1,0){5.00}}
\put(80.00,10.00){\vector(-1,0){5.00}}
\put(100.00,20.00){\vector(1,0){5.00}}
\put(110.00,10.00){\vector(-1,0){5.00}}
\put(45.00,25.00){\makebox(0,0)[cc]{$q^{12}$}}
\put(45.00,4.00){\makebox(0,0)[cc]{$q^{-12}$}}
\put(75.00,4.00){\makebox(0,0)[cc]{$q^{-2}$}}
\put(75.00,25.00){\makebox(0,0)[cc]{$q^{2}$}}
\put(105.00,4.00){\makebox(0,0)[cc]{$q^{8}$}}
\put(105.00,25.00){\makebox(0,0)[cc]{$q^{-8}$}}
\end{picture}
\caption{The tensor product graph for the $7\otimes 7$ of \gq.
\label{g2seven}}
\end{figure}

\vskip.1in
\noindent (ii) \gq:\\
The second example from section \ref{Example}, the $14+1=15$-dimensional
irrep of \gqh{} can be derived from the tensor product
\beq
7\otimes 7=27\bigoplus 14\bigoplus 7\bigoplus 1
\eeq
whose associated tensor graph is shown in figure \ref{g2seven}.

Again the tensor product graph is simply two-way connected and is therefore
equal to the generic reducibility graph.
The value of $a$ at which an edge vanishes from the reducibility graph
are given in the figure. They were
determined from $C(1)=0, C(14)=24, C(27)=28, C(7)=12$.
At $a=q^{2}$ it truncates to the reducibility graph of figure \ref{redgraph},
which describes the $14+1=15$-dimensional representation of \reff{rep14},
\reff{e015}. Various other irreducible \gqh-modules can be read off
directly from the other possible truncations of the graph.

\sect{Discussion\label{Discussion}}

In this paper we have described a practical procedure for constructing
finite dimensional representations of quantum affine algebras \ughd.
This construction relies on the reduction of tensor product
representations. We have introduced the concept of a reducibility graph
which encodes the information about which irreducible $U_q({\cal G})$-modules
have to be taken together in order to obtain an irreducible \ugh-module.
In practice we exploit the relation of the reducibility graph to
another graph, the tensor product graph, which can be constructed by
elementary means of classical representation theory.

The construction in the above sections can be extended to the case
of the tensor product $V_\l\otimes V_{\l'}$ with $\l\neq\l'$. In this case
we may still draw the associated tensor product graph,
using similar rules to the ones illustrated above; such a graph truncates
at some value of $a$, although the exact form of $a$ is not necessarily
given by (\ref{a}) and needs to be determined.
Also the construction can be applied to the tensor product of
those irreducible \ughd-modules which are reducible as $U_q({\cal G})$-modules.
Then the reducibility graph is not two-way connected even for generic $a$.
By using these methods we hope to arrive at a classification of
all finite dimensional represenations of \ughd.
All those directions
and other related aspects are under hard investigations \cite{BDGZ}.

As mentioned in the introduction, given any two finite dimensional
representations  of \ugh one can write down a spectral parameter
dependent R-matrix.
One method of doing this, applied in \cite{Zha93}, is to
insert the matrix forms of the generators in the particular
representations into the formula for the universal R-matrix.
The advantage of this method is that it is totally irrelevant whether
the representation is
reducible or irreducible, whether the tensor product decomposition is
multiplicity-free or with finite multiplicity, (the tensor product
decomposition of reducible representation with itself is always with
finite multiplicity,) and whether the representations being tensored are
the same or different. The disadvantage is, however, that this method
requires the explicit form of the universal R-matrix,
given in \cite{Kho92}, and of $e_0~ (f_0)$.

Because of this relation between the existence of a representation
of \ugh on particular \ugd-modules and the existence of the spectral
parameter dependent R-matrices for those modules, our work is related
to many works on the construction of R-matrices. In many of these
works it has been noticed that often R-matrices can only be
constructed on sums of several irreducible \ugd-modules. Our interpretation
of these observations is that only those sums of \ugd-modules carry
representations of \ughd.

The problem of constructing finite dimensional representations also
exists for the Yangians $Y(\G)$, which give the rational solutions to
the Yang-Baxter equation. Already the first paper on the problem
\cite{Dri85} by Drinfeld adresses the problem. Drinfeld is able to
give a sufficient but not necessary condition for determining wether
an irreducible $\G$-modules can carry representations of $Y(\G)$.
Later he introduced a different realization of Yangians \cite{Dri88}
in order to facilitate the construction of finite dimensional
representations, but to our knowledge, also in this realization the
problem has not yet been completely solved. The relation to our paper
lies in the fact that the rational R-matrices of the Yangian $Y(\G)$ can
be obtained from the trigonometric R-matrices of the quantum affine
algebra \ugh in a limit and therefore all $\G$-modules which we determine
to carry representations of \ugh should also carry representations of
$Y(\G)$.

Our physical motivation for studying the finite dimensional irreducible
representations of quantum affine algebras comes from the study of the
solitons in affine Toda quantum field theory. Let us explain briefly:

It is well-known that associated to every affine Lie algebra
$\hat{\G}$ there is a 1+1
dimensional affine Toda field theory, denoted $T(\hat{\G})$.
It is described by the field equations
\begin{equation}\label{todaeq}
\Box\vec{\phi}=\frac{\sqrt{-1}}{\beta}\sum^r_{i=0}n_i\vec{\alpha}_i\,
e^{\sqrt{-1}\beta\,
  \vec{\alpha}_i\cdot \vec{\phi}}\label{toda-e}
\end{equation}
$\beta$ is the coupling constant
and the $\alpha_i$ are the simple roots, $\a_0=-\sum_{i=1}^r n_i\a_i$.
For $\hat{\G}=A_1^{(1)}$ eq.(\ref{todaeq}) specializes
to the sine-Gordon (or affine Liouville) equation.

The affine Toda theory $T(\hat{\G})$ posesses symmetry generators
$e_i\,,\,f_i\,,\,h_i\,,~i=0,1,\cdots, r$, which generate the quantum affine
algebra $U_q(\hat{\G}^\prime)$
\cite{Ber91}. Here $\hat{\G}^\prime$ is the dual Lie algebra to $\hat{\G}$,
i.e., it is obtained by interchanging the roles of the roots and the  coroots.
The deformation parameter $q$ is determined by the coupling constant as
$q=e^{-\sqrt{-1}\pi/{\beta^2}}$. The central charge is zero.

The field equations \reff{todaeq} have soliton solutions. There exists a
very elegant construction of these solitons using the representation
theory of the affine Lie algebra $\hat{\G^\prime}$ \cite{Oli93}.
The solitons are found
to arrange in multiplets given by the fundamental representations of
${\G}^\prime$ (representations with a fundamental weight as highest
weight).

In the quantum theory the classical soliton solutions give rise to
particle states and we are interested in the properties of these quantum
solitons. Related work for $\hat{\G}=a_n^{(1)}$ has been done
by Hollowood \cite{Hol}.
The quantum solitons have to transform in finite dimensional multiplets
of the symmetry algebra $U_q(\hat{\G}^\prime)$.
This paper can be seen as providing some of the
necessary mathematical knowledge for extending the elegant group
theoretic understanding of the classical solitons to the quantum level.
An immediate outcome is that there are often more quantum solitons than
the classical solitons filling the fundamental representations. We
saw a concrete example: The solitons transforming in the second
fundamental representation of \gq{} (the 14-dimensional representation)
have to be completed by an additional soliton to make up the
15-dimensional multiplet of \gqh{} described by \reff{rep14} and
\reff{e015}.

\begin{center}
{\bf Acknowledgements:}
\end{center}
We thank Anthony J. Bracken and Mark D. Gould for
interesting discussions. G.W.D. also thanks them for their
hospitality during his visit at the University of Queensland.
Y.Z.Z. has been financially supported by
Australian Research Council.

%\newpage

\end{document}